\documentstyle[12pt]{article} 
\newcommand {\be} {\begin{eqnarray}} 
\newcommand {\ee} {\end{eqnarray}}  
\input psfig.sty
\begin{document} 
\pagestyle{empty}
\Huge{\noindent{Istituto\\Nazionale\\Fisica\\Nucleare}}

\vspace{-3.9cm}

\Large{\rightline{Sezione di ROMA}}
\normalsize{}
\rightline{Piazzale Aldo  Moro, 2}
\rightline{I-00185 Roma, Italy}

\vspace{0.65cm}

\rightline{INFN-1264/99}
\rightline{September 1999}

\vspace{1.cm}

\begin{center}{\large\bf Deuteron Magnetic and Quadrupole Moments with a
Poincar\'e Covariant Current Operator in the Front-Form Dynamics}
\end{center}
\vskip 1em
\begin{center} F.M. Lev$^a$,  E. Pace$^b$ and G. Salm\`e$^c$ 
\end{center}

\noindent{$^a$\it Laboratory of Nuclear Problems, Joint Institute
for Nuclear Research, Dubna, Moscow region 141980, Russia}

\noindent{$^b$\it Dipartimento di Fisica, Universit\`a di Roma
"Tor Vergata", and Istituto Nazionale di Fisica Nucleare, Sezione
Tor Vergata, Via della Ricerca Scientifica 1, I-00133, Rome,
Italy}

\noindent{$^c$\it Istituto Nazionale di Fisica Nucleare, Sezione
di  Roma I, P.le A. Moro 2, I-00185 Rome, Italy}

\vspace{2.cm}

\begin{abstract}
The deuteron magnetic and quadrupole moments are unambiguosly determined
 within
the front-form Hamiltonian dynamics, by using  a new current operator which
fulfills Poincar\'e, parity and time reversal covariance, together with 
hermiticity and the continuity equation. For both quantities the usual
disagreement between theoretical and experimental results is largely removed.
\end{abstract}
 \vspace{6.5cm}
\hrule width5cm
\vspace{.2cm}
\noindent{\normalsize{ 
To appear in  {\bf Phys. Rev. Lett.}}} 

\newpage

\pagestyle{plain}




\small

The deuteron is a good system for a test of relativistic approaches
 devoted to the investigation of hadron electromagnetic
(em) properties (see, e.g., \cite{Kondra,Tjon,Gross,Carbo,Kapta} and
Refs. quoted therein) and in particular of the accuracy of the one-body 
impulse
approximation (IA) for the current operator. It is usually believed that
effects beyond IA, e.g. meson-exchange currents,
$N\bar{N}$-pair creation terms ($Z$ graphs), and isobar configurations 
in the deuteron 
wave function are
important for the explanation of existing data. However, these effects 
are
essentially model dependent  \cite{Schia} and, furthermore, obviously 
depend on
the reference frame (see, e.g., Refs. \cite{CCKP,LPS}). 

A widely adopted framework for relativistic investigations of deuteron 
em properties is the
front-form Hamiltonian dynamics (FFHD) 
\cite{Dir,Ter}, where only the two-nucleon state is
considered and the wavefunction of the system factorizes, for any front-form 
boost,
in an eigenfunction of the total momentum times an intrinsic wavefunction. 
Thus, deuteron em form
factors, determined from three independent matrix elements 
of the  current, are given in terms of elastic em nucleon form factors 
and deuteron 
internal wave
function. In the FFHD, the one-body approximation was usually applied  
only to the relevant
 matrix elements of the 
{\em plus} component of the current in the reference 
frame
where $q^+=0$ ($q$ is the momentum transfer), while the other ones
 were properly 
defined in order to fulfill all the required properties (see, e.g.,
\cite{Kondra,CCKP,GrKo}).
In this letter, we will consider the Breit reference frame where the 
three-momentum transfer is along the spin quantization axis, which allows 
one to exploit the symmetry of the problem and to calculate 
all the non-vanishing matrix elements of the current by the same rules.

Following Ref.\cite{LPS}, let us consider the current which 
 in the Breit frame where $ \vec{P}_{\bot} = \vec{q}_{\bot}=0$ has 
the form 
\begin{eqnarray}
&&j^{\mu}(K\vec{e}_z)=\frac{1}{2}\{{\cal{J}}^{\mu} (K\vec{e}_z) +
L^{\mu}_{\nu}[r_x(-\pi)]~exp(\imath \pi S_x){\cal{J}}^{\nu}(K\vec{e}_z)^* 
exp(-\imath \pi S_x)  \}  \nonumber\\
&&{\cal{J}}^{+}(K\vec{e}_z) = {\cal{J}}^{-}(K\vec{e}_z) = 
\langle \vec{P}_{\bot} = 0, P^{'+}|\Pi J_{free}^{+}(0) \Pi| \vec{P}_{\bot} = 0,
P^{+} \rangle \nonumber\\   
&&{\vec{\cal{J}}}_{\bot}(K\vec{e}_z) = 
\langle \vec{P}_{\bot} = 0, P^{'+}|\Pi \vec{J}_{free \bot} (0) \Pi |
\vec{P}_{\bot} = 0, P^{+} \rangle .  
\label{95}  
\end{eqnarray} 
 
 In Eq. (\ref{95}),  $\Pi$ is the projector onto the subspace of 
deuteron bound states 
$|\chi_1 \rangle$ of mass $m_d$ and spin $1$, $J_{free}^{\mu}(0) = J_p^{\mu}(0) 
+ J_n^{\mu}(0)$ is the  
 one-body current, $| \vec{P}_{\bot}, P^{+} \rangle$ is an eigenstate
of the total  deuteron momentum, $ \vec{P}_{\bot}\equiv (P_x,P_y)$, 
$P^{+} \equiv (P_0 + P_z)/ \sqrt{2} = {1 \over \sqrt{2}} [(m^2_d+K^2)^{1/2} - K
]$, $P^{'+}=\frac{1}{\sqrt{2}} [(m^2_d+K^2)^{1/2} + K]$, 
$K=Q/2$, $Q^2 = -q_{\mu}^2$ and $q = P' -
P$; $L[r_x(-\pi)]$ is the element of the Lorentz group
corresponding to a rotation of $-\pi$ around the $x$ axis, $S_x$ is the $x$
component of the front-form spin operator, and $^*$ means Hermitian conjugation
in internal space. From Eq. (\ref{95}), one can obtain the expression of the
 current in any other reference frame by applying the proper 
 transformations (see, e.g., \cite{LPS}). This current operator fulfills
extended Poincar\'e covariance, hermiticity and the {\em charge normalization}, 
as well as current
  conservation  \cite{LPS}. 
The second term in the first line of Eq. (\ref{95}), which ensures hermiticity,
introduces two-body terms in the current, because of the presence of $S_x$ 
(see below).

 A relevant result of our approach is that the
extraction of elastic em form factors is
 no more plagued by the ambiguities,
related to the so called "angular condition", which are present when the free
current is used in the $q^+=0$ frame (see, e.g.,
\cite{GrKo,CCKP,KS}). In this case, one has four independent matrix 
elements of the current, while the em form factors are three \cite{GrKo}. 
Differently, in our model (Eq. (\ref{95})),  it turns out \cite{LPS} 
that only {\em three matrix elements}
$j^{\mu}_{S_z',S_z}= \langle m_d 1 S_z'| j^{\mu}(K\vec{e}_z) | m_d 1 S_z
\rangle$ are independent (e.g., $j^{+}_{0,0}$, $j^{+}_{1,1}$, and
$j^{x}_{1,0}$). Therefore, there is no longer any freedom in the construction
of the three em form factors. These matrix elements (as well as any 
other one) can be easily 
obtained by Eq. (\ref{95}) in terms 
of  the matrix
elements ${\cal{J}}^{\mu}_{S_z',S_z} = \langle m_d 1 S_z'|
{\cal{J}}^{\mu}(K\vec{e}_z) | m_d 1 S_z \rangle$.
Indeed, by using  the properties of the Wigner D-functions, one can show 
that the two terms in the first line of Eq. (\ref{95}) are 
equal for the {\em plus} component
($j^{+}_{0,0(1,1)}={\cal{J}}^{+}_{0,0(1,1)}$), while for the {\em x} component
they yield ${\cal{J}}^{x}_{1,0}/2$ and $- {\cal{J}}^{x}_{0,1}/2$, respectively
($j^{x}_{1,0}=[{\cal{J}}^{x}_{1,0} - {\cal{J}}^{x}_{0,1}]/2$)  \cite{LPS}.

 As a test of 
our current, we evaluate  the deuteron
form factors at $Q^2  = 0$, namely the magnetic moment, $\mu_d$, and
the quadrupole moment, $Q_d$, which are not affected by the
uncertainties in the knowledge of the neutron em form factors at finite
 momentum
transfers. The deuteron  moments are a longstanding
problem in nuclear physics, since it was not possible to reconcile in
a coherent approach  theoretical and  experimental values for both
quantities at the same time, by  changing the tensor content of the
nucleon-nucleon ($N-N$) interaction, or considering two-body current
contributions, both in non-relativistic and in relativistic frameworks
\cite{Tjon,Lomon,AV18,Kapta}.
Our preliminary results for the deuteron form factors at $Q^2 \neq 0$
 can be
found in \cite{LPS2}.

By using the properties \cite{LPS} of the matrix
elements of  $j^{\mu}_{S_z',S_z}$  the deuteron
form factors  can be written in terms of the matrix
elements ${\cal{J}}^{\mu}_{S_z',S_z}$ \cite{LPS2}. Then, the
 magnetic moment, in nuclear magnetons, is given by   
\begin{equation}
\mu _d = \frac{(m_p\sqrt{2})}{ m_d} \lim_{Q \rightarrow 0}   \frac{1}{ Q}
{[{\cal{J}}^{x}_{1,0} - {\cal{J}}^{x}_{0,1}] \over 2 }  ,
\label{99'}
\end{equation}
while the  quadrupole moment is
\begin{eqnarray}
 Q_d = \frac{\sqrt{2}}{ m_d} \lim_{Q \rightarrow 0}  \frac{1}{ Q^2}  
[ {\cal{J}}^{+}_{0,0} - {\cal{J}}^{+}_{1,1} ] . 
\label{101}
\end{eqnarray}

 If one adopts the free-current in the  $q^+=0$ frame, 
the angular condition is satisfied at the first order in $Q$, but it is 
violated
at the second order, for $Q^2 \rightarrow 0$ \cite{Kondra}. Therefore the 
angular condition is not a problem for the
calculation of $\mu_d$, while the quadrupole moment is not uniquely 
determined.
 Differently, following our model, both of them are well determined.

The matrix elements ${\cal{J}}^{\mu}_{S_z',S_z}$ can be easily calculated,
 by
using the action of the free current on a two-body state 
$| \vec{P}_{\bot}, P^+ \rangle | \chi_{S,S_z} \rangle$ \cite{LPS1}:  
\begin{eqnarray} 
&&\langle p_1', p_2'; \sigma_1',\sigma_2' |
J_{free}^{\mu}(0) | \vec{P}_{\bot} = 0, P^+  \rangle | \chi_{S,S_z} \rangle =
\sum_{\sigma_1"} \bar{w}(p_1',\sigma_1')[2m (f_e^{is}((p_1'-p_1)^2) -
\nonumber\\  
&&f_m^{is}((p_1'-p_1)^2) \frac{(p_1+p_1')^{\mu}} {(p_1+p_1')^2} +
f_m^{is}((p_1'-p_1)^2) \gamma^{\mu}] w(p_1,\sigma_1") 
\langle {\vec k},\sigma_1",\sigma_2' | \chi_{S,S_z} \rangle  \frac{1} {\xi}  
\label{18} 
\end{eqnarray}
where $w(p,\sigma)$ is the front-form Dirac spinor \cite{LPS1}, while $f_e^{is}$
and $f_m^{is}$ are the isoscalar electric and magnetic Sachs form 
factors of the nucleon.
The relations between the internal (${\vec k}_{\bot}$, $k_z$) and individual
nucleon variables in our reference frame are given by 
\begin{eqnarray}
&&{\vec p}_{1\bot}={\vec p}_{1\bot}'={\vec k}_{\bot},\quad p_1^+=\xi
P^{+},\quad k_z = \omega(k) (2 \xi - 1),\quad \xi' = 1 + (\xi - 1) P^{+}/
P^{'+} 
\label{19}
\end{eqnarray}
where $\omega(k) = \sqrt{m^2 + k^2}$, with $m = (m_p + m_n)/2$ the nucleon 
mass,
and $k=|{\vec k}|$. Nucleon form factors cannot be
factorized in the current matrix elements, since from Eq. (\ref{19}) one has
$(p_1'-p_1)^2 = -4Q^2 (m^2+{\vec k}_{\bot}^2)/(4 m_d^2 \xi\xi')$.

In FFHD, the internal deuteron wave function with polarization vector ${\vec
e_{S_z}}$ is given by (cf. \cite{CCKP})
\begin{eqnarray}
\langle {\vec k} | \chi_{1,S_z} \rangle = (2\pi)^{3/2} [\omega (k)/2]^{1/2} 
v({\vec k})^{-1} v(-{\vec k})^{-1} 
[\varphi_0(k)\delta_{ij} - \frac{1}{\sqrt{2}}(\delta_{ij}-
\frac{3k_ik_j}{k^2})\varphi_2(k)]
\sigma_i \sigma_y {(e_{S_z})}_j 
\label{26}
\end{eqnarray}
where a sum over the repeated indices $i,j=1,2,3$ is assumed,
 $v({\vec k})$ is the generalized Melosh matrix \cite{LPS} and $\sigma_i$ are
  the
Pauli matrices. The wave functions $\varphi_0(k)$ and
$\varphi_2(k)$ coincide with the non-relativistic $S$ and $D$ waves in momentum
representation \cite{Coester} and are normalized so that $\int\nolimits
[\varphi_0(k)^2+\varphi_2(k)^2]d^3{\vec k}=1$.

 Our FFHD results corresponding to different $N-N$ interactions are compared 
 in Table
\ref{table2} with the non-relativistic ones (for overcoming numerical
instabilities a careful analytical reduction of Eqs. (\ref{99'},\ref{101}) is
needed). The standard non-relativistic results obtained with a one-body current
crucially depend on the asymptotic normalization ratio $\eta$ of D and S wave
functions and on the $D-$state percentage, $P_D$,  but one
cannot obtain at the same time the experimental values for both $\mu_d$ and
$Q_d$.  In our Poincar\'e covariant calculation the relativistic corrections 
(RC) bring both
$\mu_d$ and $Q_d$ closer to the experimental values, except for the 
charge-dependent Bonn interaction. In Ref. \cite{CCKP},  RC have been 
calculated within FFHD by using the free current  in the $q^+=0$  frame 
and they resulted to be
 very small for $Q_d$, while
for $\mu_d$ were able to explain only part of the disagreement with
 the experimental
value. It should be stressed that
our current operator and the one used in Ref. \cite{CCKP} are different, since
both of them are obtained from the free one, but in different reference frames,
related by an interaction dependent rotation.

In Fig. 1, $\mu_d$  and $Q_d$ are reported against the asymptotic
normalization ratio, $\eta$. As already observed for the non-relativistic
calculations of $Q_d$ \cite{Rosa,Klar}, a remarkable linear behaviour appears
 for
both quantities, except for the Bonn interaction. The values 
of
$\mu_d$ and $Q_d$, suggested by this linear behaviour in correspondence of 
 $\eta^{exp}= 0.0256$, differ from the experimental ones only by $0.5\%$
and $2\%$, respectively, i.e. much less than for the non-relativistic results.
The RC to $\mu_d$ are rather large
and the total result is greater than $\mu_d^{exp}$.
This shows that, within
our framework, even the sign of explicit contributions of two-body currents is
different from the one needed in the non-relativistic case.

In summary, our results for $\mu_d$ and $Q_d$, unambiguosly calculated by 
a Poincar\'e covariant current built up from the one-body current in the Breit
reference frame where $\vec{P}_{\bot} = \vec{q}_{\bot} = 0$, show that
the total contribution of explicit two-body currents (from
meson-exchange, Z-graphs, etc.) and isobar configurations is relatively small
at  $Q^2 = 0$.
 It should be stressed that 
 explicit two-body current contributions, considered in addition to the 
 ones already contained in Eq.(\ref{95}),   must fulfill separately the 
 constraints imposed by the 
 extended Poincar\'e covariance, hermiticity and current conservation 
 \cite{LPS}. An  evaluation in our
Breit frame of explicit two-body contributions  will be performed elsewhere.

The authors wish to thank A. Kievsky for kindly providing the
deuteron wavefunctions for RSC, Av14, and Av18 interactions and R.
Machleidt for the CD-Bonn wavefunction.

\newpage
\begin{table}
\caption{Magnetic moment (in nuclear magnetons) and quadrupole moment for the
deuteron, corresponding to different $N-N$ interactions; $\mu _d^{NR}$
and $Q_d^{NR}$ are the nonrelativistic results, $\mu _d$ (LPS) and $Q_d$ (LPS)
our present results;  $P_D$ is the $D$-state percentage, and $\eta =
A_D/A_S$ the asymptotic normalization ratio.}      
\begin{center} 
\begin{tabular}{|c|c|c|c|c|c|c|}
\hline
Interaction & $P_D$  & $\eta$ & $\mu _d^{NR}$ &
$\mu_d$ (LPS) & $Q_d^{NR}$ $(fm^{2})$  & $Q_d$ (LPS) $(fm^{2})$\\
\hline 
Exp  &  & 0.0256(4) \cite{eta} &  &   0.857406(1) \cite{Lind}&  &
0.2859(3) \cite{Rosa}\\ 
RSC \cite{RSC} & 6.47 & 0.0262   & 0.8429   & 0.8611 & 0.2796 & 0.2852 \\ 
Av14 \cite{AV14} & 6.08 & 0.0265 & 0.8451   & 0.8608 & 0.2860 & 0.2907 \\
Paris \cite{Par} & 5.77 & 0.0261 & 0.8469   & 0.8632 & 0.2793 & 0.2841 \\
Av18 \cite{AV18} & 5.76 & 0.0250  & 0.8470  & 0.8635 & 0.2696 & 0.2744 \\
Nijm93 \cite{Nij} & 5.75 & 0.0252 & 0.8470  & 0.8629 & 0.2706 & 0.2750 \\
RSC93 \cite{Nij} & 5.70 & 0.0251 & 0.8473   & 0.8637 & 0.2703 & 0.2750 \\
Nijm1 \cite{Nij} & 5.66 & 0.0253 & 0.8475   & 0.8622 & 0.2719 & 0.2758 \\
CD-Bonn \cite{CDB} & 4.83 & 0.0255 & 0.8523 & 0.8670 & 0.2696 & 0.2729 \\ 
\hline 
\end{tabular} 
\end{center} 
\label{table2} 
\end{table}
\vskip 1cm
\centerline{Figure Caption}

Fig. 1. (a) Deuteron magnetic moment, $\mu_d$,  as a function of the
asymptotic normalization ratio $\eta$, for different $N-N$
interactions.  The full dot represents the experimental values for $\mu_d$ and
$\eta$; empty triangles and diamonds correspond to the non-relativistic and
relativistic results of Table \ref{table2}, respectively, while the solid and
dashed lines are linear fits for these results. Full triangle and diamond
are the results of the CD-Bonn interaction. (b) The same as in (a), but for the
deuteron quadrupole moment, $Q_d$.
\newpage
\pagestyle{empty} 
\begin{figure}
\psfig{bbllx=8mm,bblly=176mm,bburx=0mm,bbury=266mm,file=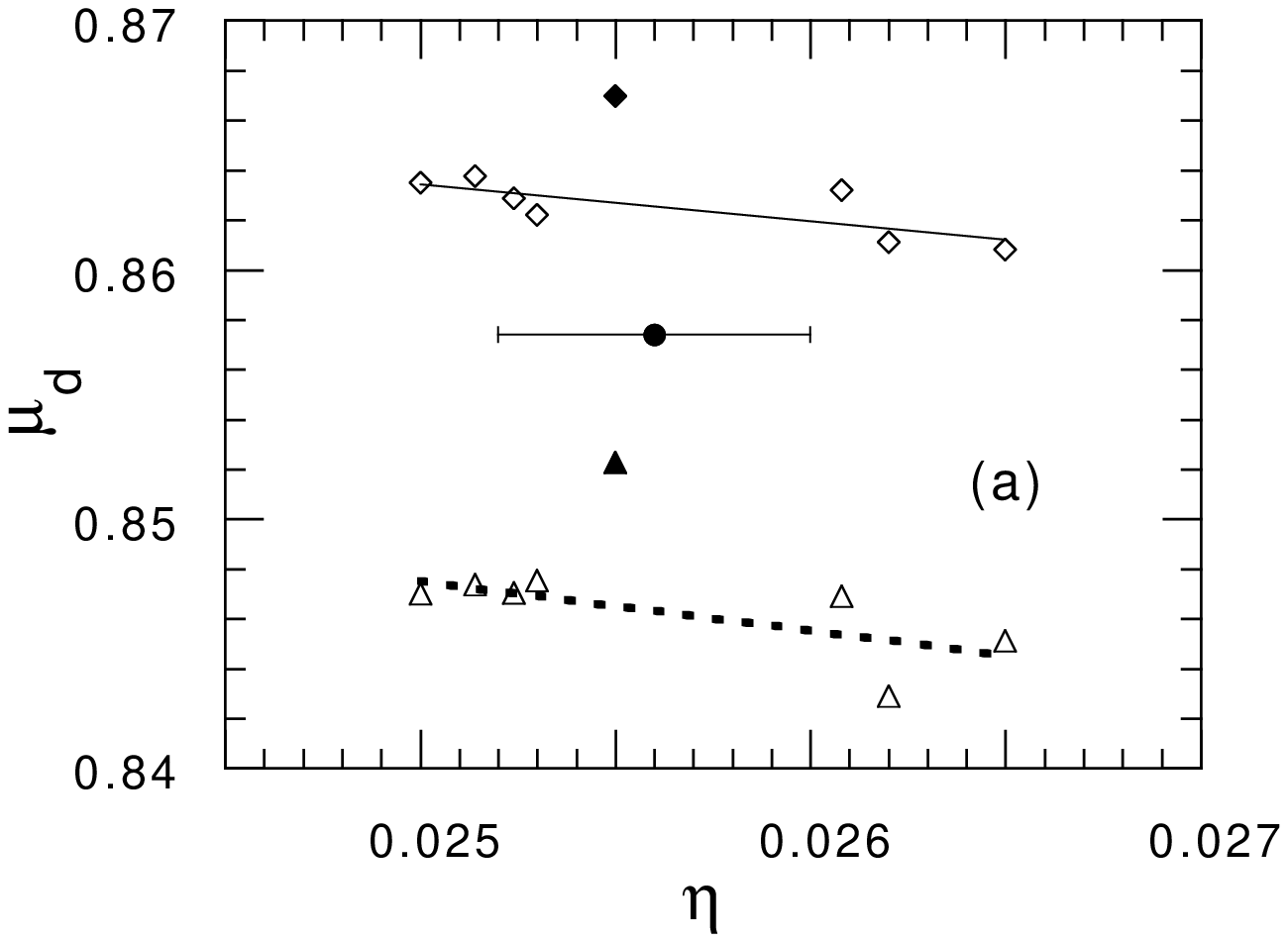}

\vspace {2cm}
\psfig{bbllx=8mm,bblly=176mm,bburx=0mm,bbury=266mm,file=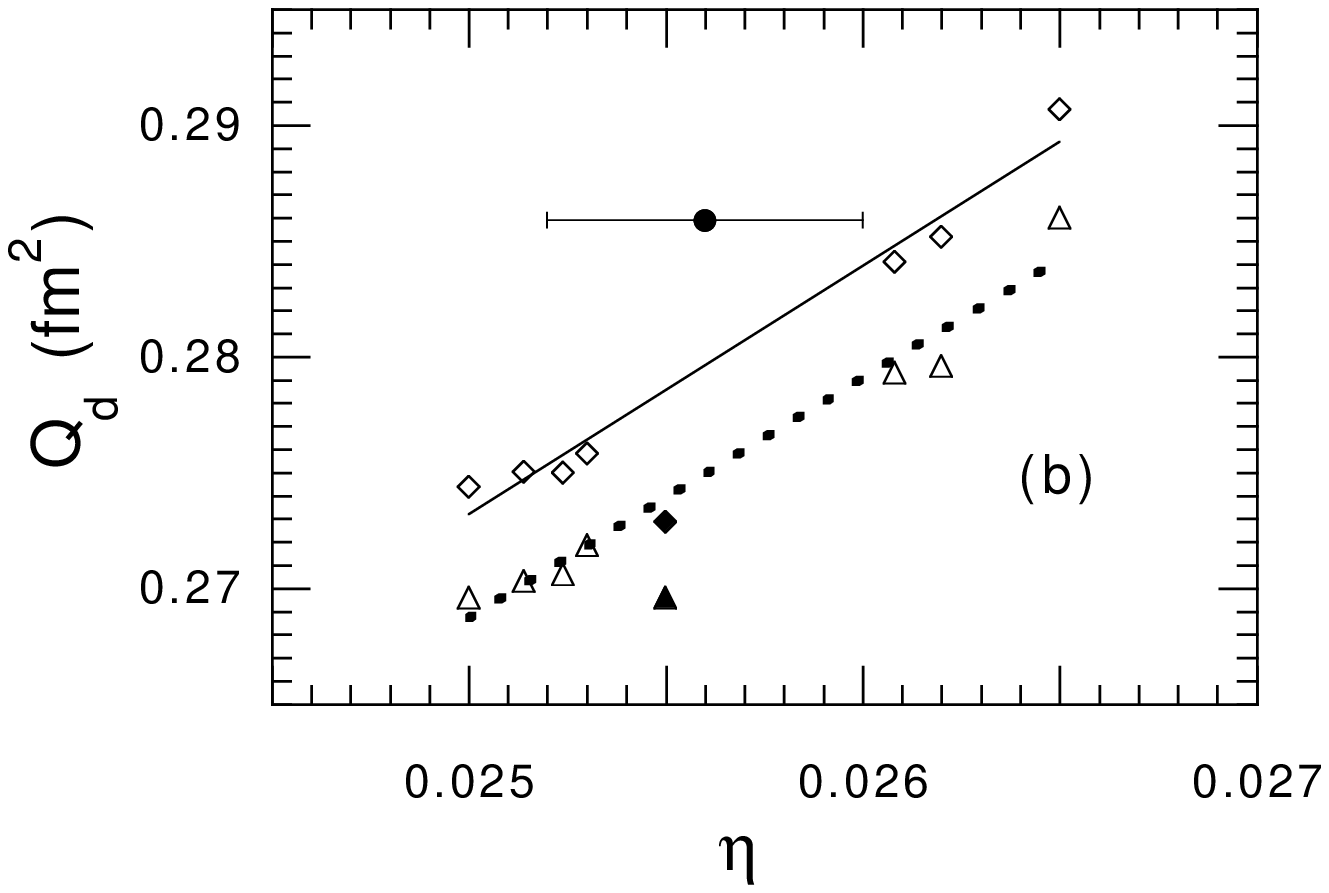}
\end{figure}

\end{document}